\documentclass[submission,copyright,creativecommons]{eptcs}
\providecommand{\event}{ThEdu 2017 Postproceedings} 
\providecommand{\thisvolume}[1]{this volume of EPTCS, Open Publishing Association}
\usepackage{underscore}           
\usepackage{paralist}       
\usepackage{graphicx}       
\usepackage{amsmath}        

\title{Prototyping \\``Systems that Explain Themselves'' \\for Education}
\author{Alan Krempler
\institute{MMM-Kommunikation\\ Graz, Austria}
\email{alan.krempler@mmm-komm.at}
\and
Walther Neuper
\institute{University of Technology\\
Graz, Austria}
\email{wneuper@ist.tugraz.at}
}
\def\titlerunning{Prototyping ``Systems that Explain Themselves''}
\def\authorrunning{A.Krempler \& W.Neuper}
\def\isac{\textit{ISAC}}
\def\sisac{\textit{ISAC}}

\begin{document}
\maketitle

\begin{abstract}
``Systems that Explain Themselves'' appears a provocative wording, in particular 
in the context of mathematics education --- it is as provocative as the idea of
 building educational software upon technology from computer theorem proving. 
In spite of recent success stories like the proofs of the Four Colour Theorem or 
the Kepler Conjecture, mechanised proof is still considered somewhat esoteric by 
mainstream mathematics.

This paper describes the process of prototyping in the \sisac{} project from a 
technical perspective. This perspective depends on two moving targets: On the 
one side the rapidly increasing power and coverage of computer theorem provers 
and their user interfaces, and on the other side potential users: What can 
students and teachers request from educational systems based on technology and 
concepts from computer theorem proving, now and then?

By the way of describing the process of prototyping the first comprehensive 
survey on the state of the \sisac{} prototype is given as a side effect, made 
precise by pointers to the code and by citation of all contributing theses.
\end{abstract}

\section{Introduction to a Never Ending Story \dots}\label{sec:intro}
\dots a story, where no end is in sight for realising specific ideas how to 
support 
learning and teaching mathematics. In the early nineties of the last century 
visionary minds were required to come up with the idea of using concepts and 
technologies from (computer) theorem proving (TP)\footnote{In this paper TP 
abbreviates the academic discipline as well as the products this discipline 
develops, proof assistants and automated provers frequently included in the 
former.} to build educational math software; these minds were Dines 
Bj{\o}rner\footnote{\url{https://en.wikipedia.org/wiki/Dines_Bjoerner}}, 
Peter Lucas\footnote{\url{https://de.wikipedia.org/wiki/Peter_Lucas_(Informatiker)}}
and Bruno 
Buchberger\footnote{\url{https://en.wikipedia.org/wiki/Bruno_Buchberger}}, who jointly 
planned such a project within the framework of UNU/IIST\footnote{\url{http://www.iist.unu.edu/}}. TP work 
done in Europe and in the US since the fifties was only known to a few 
specialists. But it was considered trustworthy enough to state the main idea for 
an R\&D project: if mathematics is the science of reasoning (and not only of 
calculating), then technology implementing such reasoning must be used as a base.
Presently there are two exceptions underpinning their educational tools 
for formal mathematics with  
TP, MathToys~\cite{url-mathtoys} and 4ferries~\cite{url-4ferries}, the former just an 
experiment and the latter already a commercial firm (tools for geometry, where
some of them also integrate TP, are not mentioned here).

Funding of the UNU/IIST project failed and respective material was used to start the 
\sisac-project. At that time computer algebra systems became readily available, 
and \sisac's decision for TP has been questioned from beginning and respective 
argumentation became more pinch-hitting over time. Taking apart Isabelle was 
much easier as is today; 
there was no Isar proof language~\cite{mw:isar07}, all commands could be 
executed on the command line and the
Emacs, driven by a specific interface~\cite{aspinall:proof-gen}, was considered insurmountable. However, numerals 
looked strange (\texttt{\#1, \#2, \dots}
)\footnote{\url{http://www.ist.tugraz.at/projects/isac/publ/mat-eng-de.pdf}}, so 
even work on reals in floating point representation and complex numbers was started and removed 
again later\footnote{\url{https://intra.ist.tugraz.at/hg/isa/rev/ab57fbfcfffd}}--- one of the several 
efforts made superfluous by Isabelle's rapid development.

In the early phases of the \sisac{} project efforts required for developing a 
usable tool have been drastically underestimated. Early field tests at technical 
schools were successful~\cite{imst-htl06,imst-htl07,imst-hpts08} in that they 
confirmed the design 
principles; but the test also showed clearly that very much work would be required 
to arrive at a system, which does not distract students from learning. 
From the very beginning much code came from students' diploma theses, master's
theses and projects. Contribution of students mostly were inspiring, and those 
from their supervisors invaluable: the latter now form an interdisciplinary 
network~\cite{url-isabelle-team} ready to constitute 
competent project teams.

Of course, there were various attempts at fund raising; most of them 
failed\footnote{\sisac's web page \url{http://www.ist.tugraz.at/isac/} reflects lack of 
funding and only serves internal information for developers.}: 
Austrian FWF rejected proposals as to far off 
basic research, SparklingScience rejected proposals because referees 
doubted that yet another computer software would improve math education; FP7 
rejected an internationally well staffed proposal, because 2/3 of the money was 
planned for development and only 1/3 for pedagogical evaluation (instead the 
other way round). The problem is that experts in engineering and mathematics 
education still don't know TP and thus cannot judge respective promises.

\bigskip
The paper roughly follows the timeline of \sisac's development. The mathematics 
engine \S\ref{sec:math-eng} was developed first; \S\ref{ssec:isab-isac} shows 
how much \sisac{} benefits from Isabelle while \S\ref{ssec:lucin} introduces an 
original contribution of \sisac{} (enabling the system to propose a next step when 
the student gets stuck). \S\ref{ssec:knowl} introduces \sisac's universe of 
mathematics knowledge and \S\ref{ssec:rewrite} explain why Isabelle's great 
simplifier is \emph{not} used. The front-end's development \S\ref{sec:front-end} 
has been started later, following a thorough design phase in 2002/2003. 
\S\ref{ssec:libisab} describes how \sisac's front-end communicates with the 
mathematics engine based on Isabelle, \S\ref{ssec:self-expl} and 
\S\ref{ssec:user-guide}  explains why \sisac{} can be called self-explanatory 
(by meeting users' expectations and by specific dialog 
guidance). Most recent requirements analysis at technical faculties led to 
specific support for a specification phase \S\ref{ssec:spec-phase}. 
\S\ref{sec:issues} collects current issues in R\&D, in particular front-end 
technologies \S\ref{ssec:front-ends} and transition to professional development 
\S\ref{ssec:professional}. The final conclusions are given in 
\S\ref{sec:conclusion}, particularly justifying the aim towards
``systems that explain themselves''.

\section{The Mathematics Engine}\label{sec:math-eng}
The mathematics engine's (abbreviated to math-engine in the sequel) repository 
has been separated from the front-end, because respective developments run 
separated after the interface in between had 
stabilised\footnote{\url{https://intra.ist.tugraz.at/hg/isa/file/2f1b2854927a/src/Tools/isac/Interpret/mathengine.sml}}.
The math-engine is designed following concepts of TP and implemented re-using 
technology from TP.

\subsection{Isabelle's Components Used in \isac}\label{ssec:isab-isac}
\sisac{} adopts as much of the TP Isabelle's~\cite{Nipkow-Paulson-Wenzel:2002} 
concepts as appropriate for engineering mathematics, and uses as much of 
Isabelle's code as well. This is in more detail:
\medskip
\begin{compactitem}
\item terms of simple typed $\lambda$-calculus, type inference and parsing of 
terms --- the basis of modelling mathematics.
\item logical contexts; it took a long time until the ``everything local'' 
principle pervaded Isabelle (and this process is still not finished); in 2004 
Isabelle's tactics started to understand contexts, \sisac{} took up this concept 
much later~\cite{mlehnf:bakk-11}. The most significant benefit for \sisac{} is, 
that formulas input by the user need not be complicated by type annotations, 
because types are inferred from the context.
\item typed matching for simplification; \sisac, however, has a specific 
simplifier for a certain reason: traces of simplification are surprisingly long, 
too long to ``explain'' to a student what is going on. But if one groups the 
rewrite rules, one can get steps fairly close to hand-written calculations, for 
instance when simplifying fractions~\cite{MG:thesis}.
\item automated provers for three tasks:
\begin{compactenum}
\item Check pre-conditions of formal specifications for problems and of guards 
for methods
\item\label{pt:derive-input} Derive a formula input by a student from the 
logical context (or reject the input); for this purpose Isabelle's provers 
combined with proof reconstruction by Metis~\cite{blanch:dis-proof-11} seem 
appropriate. Construction of most solutions for engineering problems is simple 
forward reasoning. Since this is mostly within normalising term rewriting, 
correctness of input is decidable in most cases.
\item Check post-conditions upon completion of calculations (also in sub-problems).
\end{compactenum}
Only the first two tasks are implemented; they still use only the simplifier.
\item knowledge management by theories; engineering students are \emph{not} 
expected to create \emph{new} mathematics knowledge, rather they are expected 
to use it efficiently and with understanding. For that purpose \sisac{} tries to 
support investigative access to theories; these are views on theories are 
additional to Isabelle's, see \S\ref{ssec:knowl}
\item all theories imported by multivariate analysis; this was the appropriate 
knowledge base for prototyping, which will be extended to various engineering 
disciplines. Actually, \sisac{} is able to make domains, prepared by Formal 
Methods, interactively accessible, for instance, interactions under security 
protocols by rewriting~\cite{Security-AFP}. The wide range of applications will 
enforce to re-organise theory imports (which are presently blocked by the interface, see 
\S\ref{ssec:front-ends} below)
\end{compactitem}
\medskip
The major difference between Isabelle and \sisac{} is in implementation of 
proofs versus calculations. The reason is that problem solving in engineering 
mathematics, in particular as done in academic education, is very different
from proving; we will elaborate on this in \S\ref{ssec:self-expl} below.
However, since such calculations are constructed by simple forward reasoning,
results of calculations in \sisac{} are correct by construction.

\sisac{} holds calculations in a 
\texttt{Ctree}\footnote{\scriptsize{\url{https://intra.ist.tugraz.at/hg/isa/file/2f1b2854927a/src/Tools/isac/Interpret/ctree.sml}}}
 with two kinds of nodes: (1) for steps in 
forward reasoning together with formal 
justification\footnote{\scriptsize{\url{https://intra.ist.tugraz.at/hg/isa/file/2f1b2854927a/src/Tools/isac/Interpret/ctree-basic.sml\#l41}}},
and (2) for formal 
specifications\footnote{\scriptsize{\url{https://intra.ist.tugraz.at/hg/isa/file/2f1b2854927a/src/Tools/isac/Interpret/ctree-basic.sml\#l27}}},.
The root of the tree is a specification, specifications in 
leaves hold sub-problems. The \texttt{type calcstate}\footnote{\scriptsize{\url{https://intra.ist.tugraz.at/hg/isa/file/2f1b2854927a/src/Tools/isac/Interpret/calchead.sml\#l89}}}
is given by 
a \texttt{ctree} paired with a pointer to the current position in the tree.

This design is very different from Isabelle's \texttt{structure Proof_Node} and 
\texttt{datatype state}; respective design decisions of  \sisac{} need to be 
revised when \sisac{} is going to adopt Isabelle/PIDE as discussed in \S\ref{ssec:front-ends} below.

\subsection{The Lucas-Interpreter Extends TP}\label{ssec:lucin}
In principle, there is no general method to find proofs in TP~\cite{goedel-thms}. Nevertheless, the 
development of \sisac{} has been started with the requirement that an educational system 
must model a process of problem solving such that student and system 
cooperate on equal terms: both have some knowledge (and tell it on request), 
both can check the partner's steps, both know how to do a next step --- and both 
can change roles any time (as, for instance, possible with chess software).

So in 2001 \sisac's development started with a programming language and a respective
interpreter\footnote{\scriptsize{\url{https://intra.ist.tugraz.at/hg/isa/file/2f1b2854927a/src/Tools/isac/Interpret/script.sml}}}.
At that time there was no function package~\cite{krauss:funs} in 
Isabelle, so this has been developed from scratch under supervision of Peter 
Lucas\footnote{\scriptsize{\url{https://de.wikipedia.org/wiki/Peter_Lucas_(Informatiker)}}} 
and named ``Lucas-Interpreter'' later. For instance, the program guiding user-interaction of the example shown 
in Fig.\ref{fig:isa-transparent} on p.\pageref{fig:isa-transparent} is the 
following.
{\footnotesize\label{prog:biegel}
\begin{verbatim}
    "Script Biegelinie (l::real) (q::real) v::real) (b::real=>real) (s::bool list) =" ^
    "  (let (funs::bool list) =                                                     " ^
    "             (SubProblem (Biegelinie, [vonBelastungZu, Biegelinien],           " ^
    "                          [Biegelinien, ausBelastung])                         " ^
    "                          [REAL q, REAL v]);                                   " ^
    "       (equs::bool list) =                                                     " ^
    "             (SubProblem (Biegelinie, [setzeRandbedingungen, Biegelinien],     " ^
    "                          [Biegelinien, setzeRandbedingungenEin])              " ^
    "                          [REAL l, BOOL_LIST funs, BOOL_LIST s]);              " ^
    "       (cons::bool list) =                                                     " ^
    "             (SubProblem (Biegelinie, [LINEAR, system], [no_met])              " ^
    "                          [BOOL_LIST equs, REAL_LIST [c, c_2, c_3, c_4]]);     " ^
    "       B = Take (lastI funs);                                                  " ^
    "       B = ((Substitute cons) @@                                               " ^
    "              (Rewrite_Set_Inst [(bdv, v)] make_ratpoly_in False)) B           " ^
    " in B)                                                                         " 
\end{verbatim}
}
The program text is parsed into an Isabelle term according to 
\S\ref{ssec:isab-isac}. Several syntactical details are imposed by the parser, 
for instance the capital letters of \texttt{REAL} or \texttt{LINEAR} avoiding 
name clashes with identifiers in the parser's (still global!) name space. The 
program's arguments \texttt{l\dots s} get the values from data prepared behind 
the example (no.\texttt{7.70} in the \texttt{Example browser} left in 
Fig.\ref{fig:isa-transparent} on p.\pageref{fig:isa-transparent}) checked by the precondition of a formal 
specification (shown in detail on p.\pageref{expl:spec-biegel}). The program 
body breaks down the solving process into three 
\texttt{SubProblem}s. A 
sub-problem is determined by two arguments, (1) pointers into \sisac's knowledge-base 
(see \S\ref{ssec:spec-phase} below) and (2) the list of arguments. 
Further details can be found in 
\cite{wn:lucas-interp-12,thedu16:lucin-user-view}.

The interpreter works on the program term like a debugger: at each 
tactic\footnote{\scriptsize{\url{https://intra.ist.tugraz.at/hg/isa/file/2f1b2854927a/src/Tools/isac/Interpret/tactic.sml\#l59}}}
(e.g. \texttt{SubProblem} \dots \texttt{Substitute} in the program above)
control is handed over to the 
user\footnote{\scriptsize{\url{https://intra.ist.tugraz.at/hg/isa/file/2f1b2854927a/src/Tools/isac/Interpret/script.sml\#l12}}}.
\cite{wneuper:gcd-coimbra} gives an example for
investigating programs, but this concept works for mathematics as well. The 
challenge for the Lucas-Interpreter is to find a derivation from the previous 
step and the logical context, in case the user inputs a 
step\footnote{\scriptsize{\url{https://intra.ist.tugraz.at/hg/isa/file/2f1b2854927a/src/Tools/isac/Interpret/script.sml\#l14}}}.
Derivation preserves logical consistency. For engineering 
mathematics this works surprisingly well: a method can be broken down into parts 
which are evaluated by normalising term rewriting systems.

Up to now Isabelle's function package has been well elaborated and programming 
is convenient --- and convenient programming is important for \sisac, because 
wide-spread usage can only be expected, if everyone interested in interactive 
course material can implement such content with efforts comparable with 
programming in Mathematica
. So, what is called 
\texttt{Script} above shall become\\ \texttt{partial\_function} (termination 
can\emph{not} be proven for whole classes of input data) and evaluation shall 
re-use mechanisms of Isabelle's code generator~\cite{codegen-tutorial-17}.

\subsection{Distinguished Knowledge}\label{ssec:knowl}
Software tools in formal methods, including TP, are designed to \emph{develop} 
formal models and to \emph{prove} properties of these. In contrast to that, \sisac{} is 
designed to interactively specify appropriate models from a given collection, to
make them transparent, to interactively construct solutions, and \emph{not} to develop 
new knowledge or to prove anything. Therefore \sisac{} provides a specific 
structure of knowledge and distinguishes three aspects of knowledge.

The example on p.\pageref{prog:biegel} shows how programming is related to this 
structure, particularly how\linebreak \texttt{SubProblem}s are determined by three 
different aspects. For the first

\medskip
{\small\texttt{SubProblem (Biegelinie, [vonBelastungZu, Biegelinien], 
[Biegelinien, ausBelastung])}}

\medskip\noindent
there is the \emph{theory} \texttt{Biegelinie}, the \emph{problem} referenced by 
\texttt{[vonBelastungZu, Biegelinien]} in a tree and the \emph{method} 
referenced by \texttt{[Biegelinien, ausBelastung]} in another tree. So there are 
three aspects of knowledge separated into three different data structures:
\begin{description}
\item [Theories] concern the \emph{deductive aspect} exactly as provided by 
Isabelle. Although Isabelle theories represent a directed acyclic graph (DAG), 
\sisac{} provides a tree view for the purpose of uniformity with the other 
structures. This shortcut needs to be revised as soon as respective views will 
become available in Isabelle/jEdit.
\item [Problems] concern the \emph{aspect of application}, which is represented 
by formal specifications, i.e. an assembly of input data, pre-conditions on 
these, ouput data and a post-condition relating input and output. Specifications 
are collected in a tree, which allows a simple kind of problem refinement down 
along branches, first 
applied to classes of equations for assigning the right methods 
solving specific classes~\cite{richard:da}.
\item [Methods] concern the \emph{algorithmic aspect} represented by the 
programming language introduced above by an example. A method associates a 
program with a guard, which has the same structure as a model (intoduced in \S\ref{ssec:spec-phase} below). Methods are 
preliminiarily stored in a tree, but in contrast to problems here is no 
specific reason for this data structure.
\end{description}
\isac{} uses Isabelle's theory management also for defining problems and methods. 
However, the structure of problems and methods is different from theories. For 
instance, the problem (i.e. the formal specification) of a linear equation is 
polymorphic for integers (i.e. rings), for rational, real and complex numbers 
(i.e. fields), for vector spaces, etc. Such a problem can be defined somewhere 
in Isabelle's DAG of theories as soon as a predicate \texttt{is\_linear} is 
defined and a post-condition can be formulated. 

So storing problems and methods in respective trees proceeds independently from 
theory evaluation. Originally a global reference variable stored these trees, later 
Isabelle enforced to use \\ \texttt{Unsynchronized.ref} and finally parallelisation 
of theory evaluation enforced to use the functor \\ \texttt{Theory\_Data} 
\cite{isa-impl-17} in Isabelle/ML.

\medskip
The above separation of knowledge appears to anticipate current R\&D in formal 
methods~\cite{sys-of-sys-15}: at least a distinction between definition of 
language elements (as in theories) and definition of re-usable specifications 
(as in problems) becomes apparent. Users' interactions on the three aspects of 
knowledge will be discussed in \S\ref{ssec:spec-phase} below.

\subsection{A Specific Rewrite Engine}\label{ssec:rewrite}
\begin{figure} [htb]
  \centering
  \includegraphics[width=145mm]{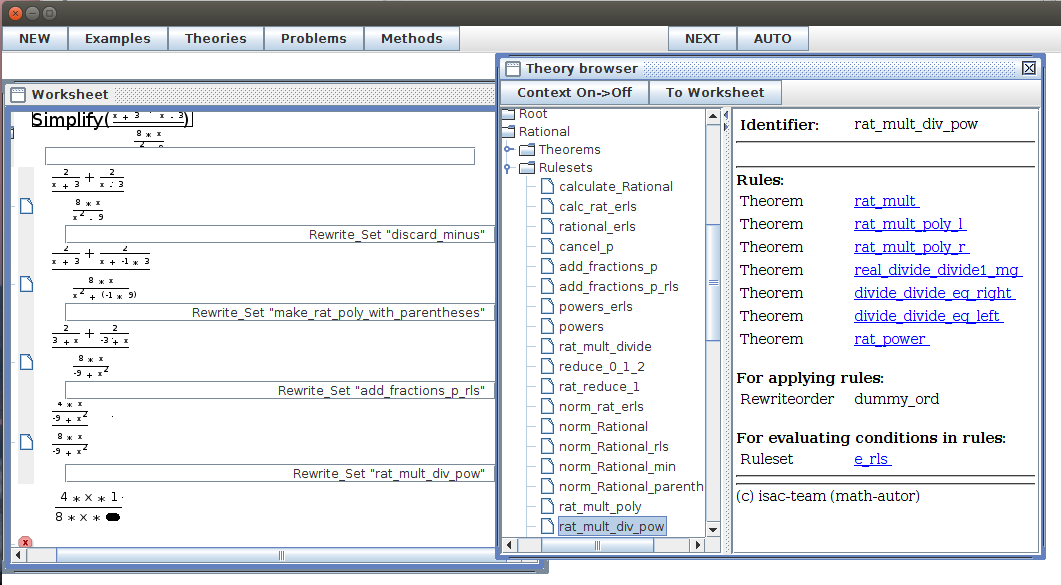}
  \caption{\sisac's close to hand-written calculations.}
  \label{fig:isac-calc}
\end{figure}

Isabelle's simplifier was a highly elaborated and complicated component already in 2000. So it 
was hard to decide, how to adapt it to \sisac's requirements. When early design 
revealed, how much Isabelle's and \sisac's requirements differ, the decision was 
for developing the simplifier from scratch and for re-using only Isabelle's 
\texttt{Pattern.match}. Presently \sisac{} requires the simplifier for three purposes.
\begin{enumerate}
\item \textbf{Evaluate programs} during Lucas-Interpretation 
(\S\ref{ssec:lucin}), in particular code concerned with list processing. 
Isabelle provides an elaborated collection of functions for this purpose, which 
need to be evaluated efficiently. \sisac{} would have gotten this for free if 
Isabelle's code generator~\cite{codegen-tutorial-17} would have existed at the
time of development.
\item \textbf{Evaluate preconditions} and postconditions of specifications after 
instantiation with the data of a particular example. \sisac{} uses Isabelle's 
$\lambda$-terms (\S\ref{ssec:isab-isac}), i.e. deep embedding of mathematical 
formulas and implementation by ML functions. Evaluation of such functions is 
alien to Isabelle's simplifier; \sisac's simplifier has to evaluate such 
functions embedded into terms with logical connectives.
\item \textbf{Simplify mathematical expressions} as close to traditional work by 
paper and pencil as possible. An example is given by Fig.\ref{fig:isac-calc}:
A fraction is simplified in collaboration between system and user; the latter 
inputs the next step (the cursor is the black spot bottom left). The user had 
asked \sisac{} for justification of steps, shown at the right margin. In the last but one step the 
Lucas-Interpreter has decided (by a program similar to the one on 
p.\pageref{prog:biegel}) to apply a group of seven theorems \texttt{rat\_mult} 
\dots \texttt{rat\_power} assembled under the identifier 
\texttt{rat\_mult\_div\_pow}. These are shown in the \texttt{Theory browser}, a 
specific view on theories (\S\ref{ssec:knowl}), in the right window of Fig.\ref{fig:isac-calc}.
\end{enumerate}
With the experiences gained by \sisac's simplifier so far the question has to be 
re-raised, whether the proprietary component would be better replaced by native 
Isabelle components, Isabelle's simplifier and evaluation machinery of the code 
generator. However, such grouping is not trivial with respect to confluence and 
termination~\cite{pl:baader98}, and compiling terminating and confluent groups 
might not preserve these two properties indispensable in rewriting. So specific 
tools like~\cite{Korp2009} might be considered to assist mathematics authors in compiling 
terminating and confluent term rewriting systems (called \texttt{Ruleset}s here).

A fundamental design idea of \sisac{} is to smoothly extend traditional customs 
and notation of engineers to a more rigorous level. The simplifier is relevant 
for this idea, because engineering problems can be broken down to sub-problems 
where most of them are just rewriting. 
And experience with simplification, equation solving, differentiation, etc has 
shown that such grouping of theorems can achieve 
traces fairly close to traditional work by paper and pencil.

\section{The Front-End}\label{sec:front-end}
The design for an \sisac-front-end was an adventure. The first attempt  
\cite{fink:da} went straight into implementation alongside the implementation of 
the math-engine --- this was a flop, the user requirements were 
completely unclear: What can users request for learning from a novel kind of 
math-engine (with features not as clearly expressed as in 
\S\ref{sec:math-eng} above)?

So in 2002 an outstanding team of four students 
\cite{RG04:thesis,AG:thesis,MH04:thesis,AK04:thesis} was formed and led by an expert\footnote{Klaus Schmaranz 
 \url{http://www.consonya.com}.}
in thorough engineering of user requirements. This resulted in a substantial 
documentation~\cite{isac:all} and a design guiding development up to the date of 
this article.
However, the documentation could not be kept up to date by students,
rather documentation is scattered across thirty diploma theses 
now\footnote{\url{http://www.ist.tugraz.at/isac/Publications_and_Theses\#Theses}},
which are also cited in this paper (thus the lengthy bibliography). Recently a 
second round of user requirements engineering started in collaboration with 
staff of engineering faculties~\cite{isac-doc2}, which is under construction and 
already led to novel ideas for support in the specification phase.

\medskip
Before turning to a description of design and implementation for the front-end 
(accompanied with a short description on expectations for users with respect to 
``systems that explain themselves'') interfaces between Isabelle's and \sisac's 
front-ends and respective back-ends are considered as follows.

\subsection{The Interface Math-Engine --- Front-End}\label{ssec:libisab}
\begin{figure} [htb]
  \centering
  \includegraphics[width=130mm]{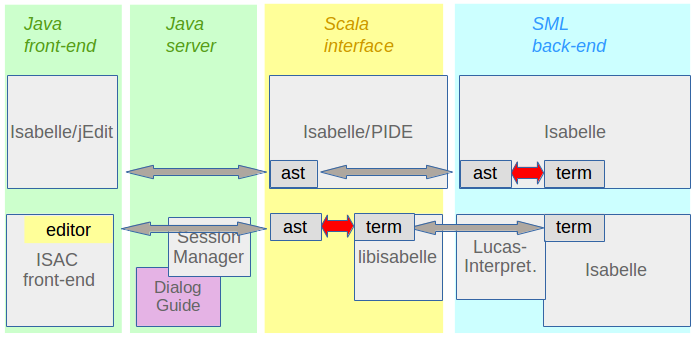}
  \caption{Interfaces of Isabelle and of \sisac.}
  \label{fig:ast-trans}
\end{figure}

The sustainable evolution of Isabelle over the years not only influenced 
\sisac's mathematics engine, but also the interface between the latter and 
\sisac's font-end. In 2000, when \sisac{} started development,  Isabelle was 
still used via tactics on the command line. So the standardised streams of Unix, 
\texttt{stdin} and \texttt{stdout} were the interface between \sisac's 
math-engine in SML and the \sisac{} 
front-end in Java. When the proof language Isar~\cite{mw:isar07} superseded 
Isabelle's tactical language, also the more advanced user-interface 
Isabelle/jEdit~\cite{makar-jedit-12} replaced the Emacs-interface. Also Isabelle/jEdit motivated a new 
kind of interface Isabelle/PIDE~\cite{DBLP:conf/itp/Wenzel14} as shown in the 
top area of Fig.\ref{fig:ast-trans} on the next page.

Isabelle/PIDE came along with dropping \texttt{stdin} and \texttt{stdout} in 
Isabelle2013-1 --- \sisac{} was in danger to lose connection between front-end
and back-end. But there was good luck: The well-known verification system Leon 
was dependent on Isabelle's \texttt{stdin} and \texttt{stdout} 
as well, so there were resources to develop the \texttt{libisabelle}~\cite{Hupel2016} interface 
as a slim version of Isabelle/PIDE and to maintain it 
reliably\footnote{\url{https://lars.hupel.info/libisabelle/}} 
--- and \sisac{} now re-uses this interface as shown in the bottom area of 
Fig.\ref{fig:ast-trans}.

Recently an experimental formula editor has been implemented in \sisac{} 
\cite{mmahringer}, already shown in the left window of Fig.\ref{fig:isac-calc}. 
This editor is not only written in Scala 
(indicated by yellow background in Fig.\ref{fig:ast-trans}), it also
raises architectural issues: An 
editor in the \sisac{} front-end uses Java Swing and the underlying 
data structure is an ``annotated syntax tree'' (AST) close to visual 
representation. However, \texttt{libisabelle} transfers Isabelle's 
$\lambda$-terms as Scala trees (which nicely integrate with Java). So 
Isabelle's translation between \texttt{terms} (appropriate for mathematics) and 
ASTs (appropriate for presentation) had to be transferred from \texttt{SML} to 
\texttt{Scala}, indicated by red arrows in Fig.\ref{fig:ast-trans}. The 
question, whether this transfer is another detour in \sisac's prototyping 
process, will be discussed at the end of the paper. \texttt{DialogGuide} and 
\texttt{SessionManager}, located on the \sisac{} server, will show up again in 
Fig.\ref{fig:dialog-guide-ak2} below and will be discussed there. The 
\texttt{Lucas-Interpreter} in the back-end has been introduced in 
\S\ref{ssec:lucin}.

\subsection{A Self-Explanatory System}\label{ssec:self-expl}
From students' perspective the main idea of \sisac{} is to support learning 
math by doing math on a \emph{model of math} implemented in software. An analogy
is good chess software which is used by champions to explore new strategies 
as well as by novices to learn how to play chess. Such software is a \emph{complete} 
(with support for the complete game from opening to final) and 
\emph{interactive} model of chess, where moves are analogous to steps in solving problems in 
engineering math. But chess software is not a \emph{transparent} model: 
watching cascades of valuation functions at work millions of times is not 
informative.

\begin{figure} [htb]
  \centering
  \includegraphics[width=150mm]{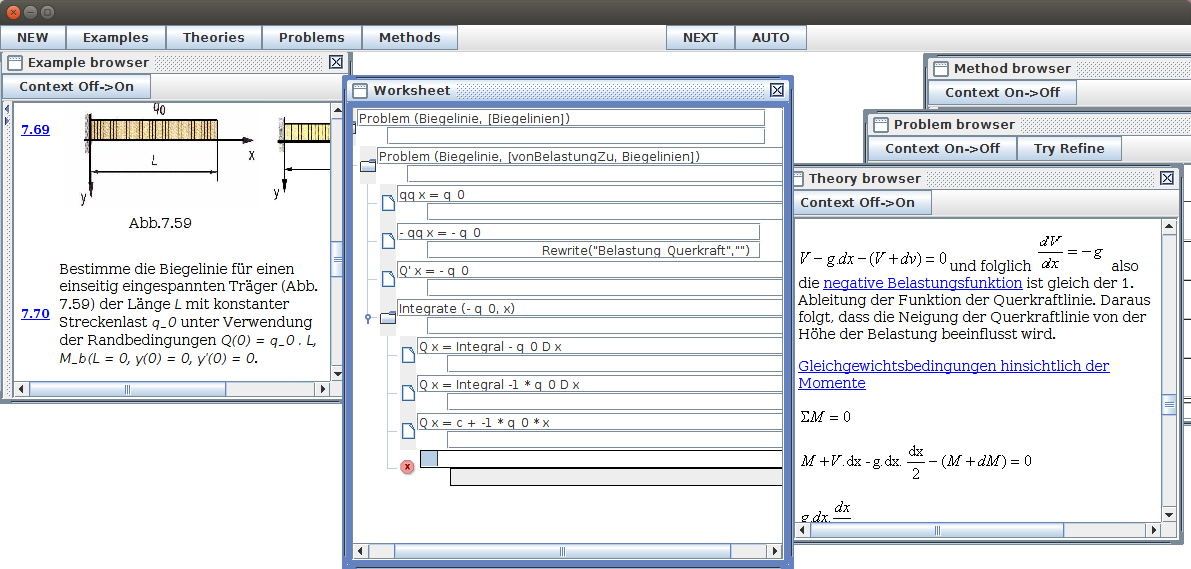}
  \caption{\sisac's front-end.}
  \label{fig:isa-transparent}
\end{figure}

So a fundamental design idea of \sisac{} is to provide
a \emph{complete, interactive and transparent 
model of math} --- the system not only explains itself at the user-interface 
(like a mobile), but also explains itself by 
exhibiting it's internal structure 
(down to a certain level):
The \sisac{} prototype is \emph{complete} in the sense that is 
supports all phases of problem solving: Fig.\ref{fig:isa-transparent} shows the 
phase of solving the \texttt{Problem} given by example \underline{7.70} in the 
electronic textbook on the left (entitled \texttt{Example browser}).
Behind the example the input data, the theory \texttt{Biegelinie} and 
suggestions of a specification \texttt{[Biegelinien]} is hidden --- design of 
interactions on these during phases of modelling and specifying will be described 
in \S\ref{ssec:spec-phase}. 

The \texttt{Worksheet} in the middle resembles a sheet of paper used for 
constructing a solution of the problem. \S\ref{ssec:rewrite} mentioned that 
\sisac{} should adapt to engineers' customs
and notations as much as possible -- which is better accomplished by the 
experimental editor in Fig.\ref{fig:isac-calc} on p.\pageref{fig:isac-calc} than 
in Fig.\ref{fig:isa-transparent}. The \texttt{Worksheet} shows the 
student at input for simplifying an integral. In addition to the functionality of 
a sheet of paper the prototype already provides various kinds of support:
\begin{compactitem}
\item The system will check the input reliably according to 
Pt.\ref{pt:derive-input} on p.\pageref{pt:derive-input}.
\item If the student gets stuck, the system can provide a \texttt{next} (the button in Fig.\ref{fig:isa-transparent}) step due to 
Lucas-Interpretation (\S\ref{ssec:lucin}); how the latter is used for 
adaptive user guidance, this will be shown in \S\ref{ssec:user-guide}.
\item In case the system does steps, the student can ask for explanation by 
requesting a derivation: in the step from $-qq\;x = -q_0$ to $Q^\prime\;x = 
-q_0$ this is only the theorem \texttt{Belastung\_Querkraft} and 
\texttt{Rewrite} is the tactic applying this theorem.
\item For each theorem respective Isabelle proofs can be inspected. However, lookup of 
underlying definitions by mouse click like in Isabelle/jEdit is not yet 
realised.
\item Each theorem can be accompanied by multimedia data with explanations; an 
example for theorem \texttt{Belastung\_Querkraft} is given in the \texttt{Theory browser} at the right.
\item The specification and the method underlying the problem solution can be 
inspected in the \texttt{Problem browser} and \texttt{Method browser} 
respectively.
\end{compactitem}
So, an interested student can find all kinds of explanations within all three 
aspects of mathematics, the deductive, applicative and algorithmic aspect 
according to \S\ref{ssec:knowl}. Field tests, performed so far 
\cite{imst-htl06,imst-htl07,imst-hpts08}, indicate that digital natives 
indeed experience such a system as self-explanatory.

\medskip
Furthermore, TP-technology offers various self-explanatory features not yet 
adopted in the prototype: For instance, Isabelle has a counterexample generator 
\cite{blanch:dis-proof-11}, or~\cite{schrein:fin-mod-17} can generate specific examples 
even for implicit definitions.

\subsection{Adaptive User Guidance}\label{ssec:user-guide}
Section \S\ref{ssec:self-expl} introduced an analogy between learning chess and learning 
mathematics by software as well as the fundamental design idea of \sisac{} to 
be a complete, interactive and transparent model of math --- a clear and 
unadorned model.
Therefore the main aim of \isac{}'s user guidance is to 
make the experience with the self-explanatory system more efficient, 
more smooth and motivating, without addressing extrinsic motivation. 
We do not go for mimicking a human tutor.

\begin{figure} [htb]
  \centering
  \includegraphics[width=120mm]{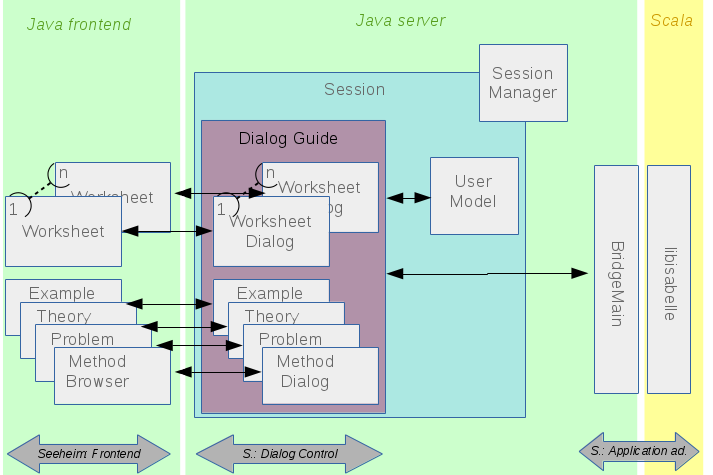}
  \caption{User interaction and its relations to code and system architecture.}
  \label{fig:dialog-guide-ak2}
\end{figure}

A \texttt{DialogGuide}\footnote{\scriptsize{\url{https://intra.ist.tugraz.at/hg/isac/file/b2fd3773f54b/isac-java/src/java/isac/session/DialogGuide.java}}}
is responsible for implementing the requirements 
regarding processes of problem solving introduced in the first paragraph of \ref{ssec:lucin}. 
In doing so, it implements the Dialog Control component of the Seeheim Model~\cite{seeheim}, 
which was chosen as a template for \isac{}'s user interaction architecture.
The structure of the \texttt{DialogGuide}, its operating infrastructure and the relation of the architecture 
to the Seeheim Model are illustrated in Fig.\ref{fig:dialog-guide-ak2}; the 
\texttt{DialogGuide} has already be mentioned in Fig.\ref{fig:ast-trans} on p.\pageref{fig:ast-trans}.
The left part of Fig.\ref{fig:dialog-guide-ak2} depicts all GUI elements 
introduced in Fig\ref{fig:isa-transparent} on p.\pageref{fig:isa-transparent}: 
\texttt{Worksheet}, \texttt{ExampleBrowser}, \texttt{TheoryBrowser}, 
\texttt{ProblemBrowser} and \texttt{MethodBrowser}; they correspond to the Frontend component of the Seeheim Model. In the central part, the \texttt{SessionManager} 
assigns one \texttt{Session} per user which contains a \texttt{DialogGuide} proper and a 
\texttt{UserModel}\footnote{\scriptsize{\url{https://intra.ist.tugraz.at/hg/isac/file/b2fd3773f54b/isac-java/src/java/isac/users/UserModel.java}}}.
The \texttt{DialogGuide} coordinates a couple of dialogs: one 
\texttt{WorksheetDialog}\footnote{\scriptsize{\url{https://intra.ist.tugraz.at/hg/isac/file/b2fd3773f54b/isac-java/src/java/isac/wsdialog/WorksheetDialog.java}}}.
per \texttt{Worksheet} (so several (variants of) calculations can be done in 
parallel) and one 
\texttt{*Dialog}\footnote{\scriptsize{\url{https://intra.ist.tugraz.at/hg/isac/file/b2fd3773f54b/isac-java/src/java/isac/browserdialog/BrowserDialog.java}}}.
per \texttt{*Browser}. Note that \emph{Dialog} in this context does not denote 
widgets on screen, but an internal component moderating the flow of user interaction.


The \texttt{UserModel} represents the system's knowledge about a user
which can be used by the \\ \texttt{DialogGuide} to adapt its behaviour
towards this specific user.  As finding the relevant parameters for a
\texttt{UserModel} is subject to further interdisciplinary research, a
testbed under the name of
\texttt{UserLogger}\footnote{\scriptsize{\url{https://intra.ist.tugraz.at/hg/isac/file/b2fd3773f54b/isac-java/src/java/isac/users/UserLogger.java}}}
has been developed~\cite{fkober-bakk} to assist in research and
prototyping.

On the far right, lastly, \texttt{BridgeMain} and \texttt{libisabelle} connect to the math-engine and constitute the Application adapter of the Seeheim Model.

In order to prepare the \texttt{DialogGuide} for meeting the challenges of 
expected complexity, user interactions are not determined by Java 
code\footnote{\scriptsize{\url{https://intra.ist.tugraz.at/hg/isac/file/b2fd3773f54b/isac-java/src/java/isac/wsdialog/WorksheetDialog.java\#l699}}},
but by a rule-based system~\cite{mkienl-bakk}.

In the prototype implementation of this architecture, the \texttt{DialogGuide} does not
interfere with the flow of communication as the \texttt{UserModel} does not yet differentiate users,
both awaiting a major development effort in collaboration with 
experts in user modelling and with experts in psychology of mathematics 
education.

Anticipating such collaboration conceivable interventions of the 
\texttt{DialogGuide} include:
\begin{description}
 \item [Sharing knowledge] in the sense of communicating steps in a calculation
 done by the user or the math-engine to the other partner for
 inspection. Information about such a step taken includes the resulting formula 
 but in addition to that may include the tactics applied to connect the
 formula to the previous one.


 \item [Deciding which knowledge to share,] as not every detail known to the
 TP is necessarily of interest to every user on every occasion. This decision includes 
 the question how much knowledge to share. In addition to depending on a 
 particular user, this decision can depend on the context of the calculation, 
 eg. whether done to solve a problem, to playfully explore or to take an exam.
 \item [Deciding when to share knowledge,] which can depend on how far into a calculation
 we have proceeded as well as how much time has passed.
 \item [Asking for explanations] which is equivalent to giving reasons for
 steps taken in a calculation. In \isac{}'s architecture this would mean 
 requesting to make transparent which tactics were employed in achieving a result
\item [Asking for activity] such as selecting a method to solve the problem at hand,
 computing the next step or proposing the next tactic to apply.
 \item [Deciding what to ask for] basically involves the same questions as deciding which
 knowledge to share. With the requirement that the \texttt{DialogGuide} treat both 
 partners on equal terms \S\ref{ssec:lucin}, it is just a matter of point of view whether 
 something passed on by the  \texttt{DialogGuide} has been 
 requested by one partner or offered by the other.
 \item [Deciding when to ask] for activity or for explanations.
 \item [Choosing the right level of abstraction] by grouping rewrite-rules as mentioned in \S\ref{ssec:isab-isac} 
 or selecting specific explanatory material. 
 \item [Switching roles] of the user and the math-engine boils down to
 exchanging the actions of sharing and requesting information. Again, the \texttt{DialogGuide} 
 does not make any difference between math-engine and user.
\end{description}
Such interventions of the \texttt{DialogGuide} are supported by \isac{}'s software architecture.

Given the experience from \sisac's prototyping, collaboration with experts in 
user modelling and in psychology of mathematics education will follow these 
guidelines:

The \texttt{DialogGuide} implements its interventions by presenting only part of the 
calculation tree to the respective partners and by blocking requests or inserting
additional requests into the communication.

Adaptive user guidance requires that the above decisions be not hard-coded,
but be made to fit a specific situation as closely as possible.
How to arrive at the right decisions will depend on the user, her knowledge, 
experience and aims on one hand and on the task being tackled on the other hand.
We can assume that the specifics of the task can be derived from the problem and the 
method of the current calculation, both of which are known.
Users can be categorised on a very general level as ``beginner'' or ``expert''.
A more refined model of user expertise could take into account whether a user is ``familiar with'',
``aware of'' or ``completely new to'' specific theories or even individual rewrite rules.
Even personal traits such as endurance or resistance to stress could make for
a better learning experience if properly considered.
Finding relevant parameters for a \texttt{UserModel} is subject to research
in fields other than computer science.


Finally, the \texttt{UserModel} need not remain static. Given that web sites can 
adapt to users by observing and analysing their browsing behaviour,
the \texttt{UserModel} could get to know the user more closely from 
experience gained in past interactions.

There are high expectations in user guidance by \sisac; some ideas how to exploit the 
potential are: support for rule-application~\cite{kremp.np:assess} and handling of 
error-patterns~\cite{darocy:master}.



The architecture described above has the major benefit that
authoring mathematics (by writing programs, see \S\ref{ssec:lucin}) is
strictly separated from authoring dialogues (by using the rule-based
system~\cite{mkienl-bakk}), thus disentangling expertise often
unpleasantly mixed in authoring educational systems.

\subsection{Specific Support for a Specification Phase}\label{ssec:spec-phase}
Support for formal specification has been a concern of prototyping from the beginning up to the 
present, when usability at engineering faculties is being considered 
\cite{isac-doc2}. From the beginning it was clear that solving problems in 
engineering mathematics starts with translating observations from the real world 
into formulas. In order to support the student in this effort, \sisac{} hides 
minimal data behind each example, for instance no.\texttt{7.70} in 
Fig.\ref{fig:isa-transparent} has this hidden formalisation:

{\footnotesize\it\begin{tabbing}
\=123\=12\=12\=12\=12\=123\=123\=123\=123\=\kill
\>\>$[$\>$($\>$[$\> Traegerlaenge $L$, Streckenlast $q_0$, Biegelinie $y$,\\
\>\> \> \> \> Randbedingungen $[\,Q\,0=q_0\cdot L,\;M_b\,L=0, \;y\,0=0, 
\;\frac{d}{dx}y\,0=0]$, FunktionsVariable $x$ $\;]$\\
\>\> \> \> $($"Biegelinie", $[$"Biegelinien"$]$, 
$[$"IntegrierenUndKonstanteBestimmen2"$\;]\;)\;)\;]$
\end{tabbing}}
\noindent
The first two items are input data (symbolic in this case), the second line 
starts with (a sub-term of) the post-condition and the third line is a triple 
referencing theory, problem and method, respectively (see \S\ref{ssec:knowl}).

These data are used by \sisac{} to help students with input to the formal 
specification, which is the following for the example in 
Fig.\ref{fig:isa-transparent}:

\label{expl:spec-biegel}
{\footnotesize\begin{tabbing}
1234\=123\=12\=12\=12,\=Postcond \=: \= $\forall \,A^\prime\, u^\prime 
\,v^\prime.\,$\=\kill
\>\texttt{01}\>Problem (Biegelinie, [Biegelinien])\\
\>\texttt{02}\>\> Specification:\\
\>\texttt{03}\>\>\> Model:\\
\>\texttt{04}\>\>\>\> Given  \>: ${\it Traegerlaenge}\;L, \;{\it Streckenlast}\;q_0$  \\
\>\texttt{05}\>\>\>\> Where  \>: $q_0 \; {\it ist\_integrierbar\_auf}\;[0,L]\;\;\;\;\;\;$ \\
\>\texttt{06}\>\>\>\> Find   \>: ${\it Biegelinie}\;y$ \\
\>\texttt{07}\>\>\>\> Relate \>: ${\it Randbedingungen}\;[Q\,0=q_0\cdot L,\;M_b\,L=0, \;y\,0=0, \;\frac{d}{dx}y\,0=0]$\\
\>\texttt{08}\>\>\> References:\\
\>\texttt{09}\>\>\>\> Theory \>: Biegelinie \\
\>\texttt{10}\>\>\>x\> Problem\>: $[$"Biegelinien"$]$ \\
\>\texttt{11}\>\>\>o\> Method \>: $[$"IntegrierenUndKonstanteBestimmen2"$]$ \\
\>\texttt{12}\>\> Solution:
\end{tabbing}}
\noindent
The \texttt{Specification} consists of a \texttt{Model} and \texttt{References}, 
where the ``x'' and ``o'' indicate selection whether the model is shown for the 
\texttt{Problem} (as specification) or for the \texttt{Method} (as guard). 
Preconditions given in \texttt{Where} are checked by an automated prover 
(\S\ref{ssec:rewrite}). The post-condition, partially given by \texttt{Relate}, 
should be checked as soon as a result has been constructed (this check is not 
yet implemented in the prototype). The \texttt{Solution} is the calculation shown in the 
\texttt{Worksheet} of Fig.\ref{fig:isa-transparent} after the window for 
specification has been closed.

\begin{figure} [htb]
  \centering
  \includegraphics[width=115mm]{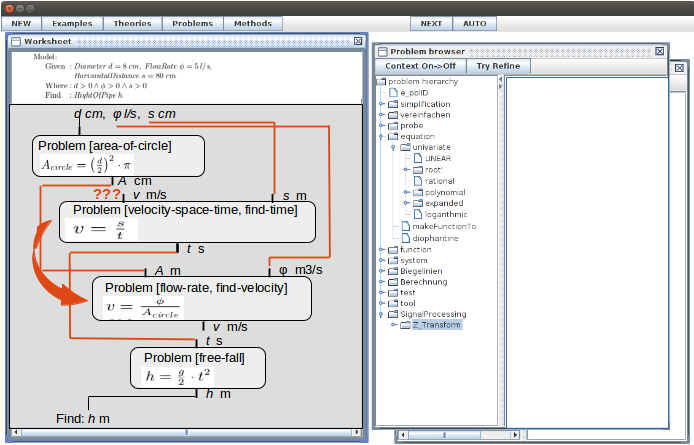} 
  \caption{Sequencing of sub-problems ahead of solve-phase.}
  \label{fig:sub-probl}
\end{figure}

Field tests showed that high-school students (and teachers!) are not interested 
in formal specification. While ``Formal Methods'' are still on the fringes in 
most engineering studies, academic staff is well aware that problem solving 
involves not only assigning methods to sub-problems (as described above) but 
also appropriate selection of problems and respective sequencing. 
Fig.\ref{fig:sub-probl} shows an example for what is in the pipeline toward 
improved support for the specification phase.
In the example the \texttt{Problem}s have already be selected from the 
\texttt{Problem browser} right and placed into a panel on the left. Here the 
problems can be moved around until a sequence is found such that each problem 
gets appropriate input from a problem above (which means that this problem has 
to be solved first). Respective operations are shown dynamically on a slide 
movie\footnote{\url{http://www.ist.tugraz.at/projects/isac/publ/movie-sub-problems.pdf}}. 
Such sequencing also would be helpful for the \texttt{SubProblem}s listed in the 
example program on p.\pageref{prog:biegel}.

This panel for selecting and sequencing sub-problems is not yet implemented in 
the prototype. However, implementation seems straight forward: Isabelle's type 
system will be helpful in matching the input and output data from sub-problems.

\section{Present Issues and Future Work}\label{sec:issues}
The never-ending process of prototyping a new TP-based generation of tools for 
engineering education has a stopover in this paper such that the next steps 
become clear. With the survey on design, on architecture and benefits for students 
given in the previous sections, specific technical issues were clarified. Before 
an account of specific tasks and estimates on respective development efforts are 
given, a big open question needs to be tackled \dots

\subsection{Where shall Front-Ends Go?}\label{ssec:front-ends}
Isabelle is well underway towards a ``Prover's Integrated Development 
Environment (PIDE)''~\cite{makar-jedit-12}. \\ PIDE's document model 
\cite{Wenzel-11:doc-orient} is powerful such that it can drive 
front-of-the-wave IDEs like MicroSoft 
VSCode\footnote{\url{http://sketis.net/2017/isabellevscode-1-0-in-isabelle2017}}, 
while connecting with HTML5 for interaction on standard browsers~\cite{Lueth2013} 
received little attention. A look ahead into the future was made in~\cite{EPTCS79.9} towards distributed version control and multi-user session 
management; however, the respective  FP7 proposal was rejected 
as mentioned in \S\ref{sec:intro}. Also ideas about collecting formal 
mathematics into wikis~\cite{Urban2010} are around for some time, but not yet 
realised.

On the other hand, \sisac's design envisaged a web-based system from the very 
beginning~\cite[p.13]{isac:all}. An intermediate experiment with launching 
\sisac's front-end via Java Webstart~\cite{tzilling:master} is promising. But 
there are good reasons to replace \sisac's proprietary front-end 
(Fig.\ref{fig:ast-trans} on p.\pageref{fig:ast-trans})
with some standard component in order to reach 
professional usability with reasonable effort. As an educational tool \sisac{} 
shall go in direction of handhelds. But respective technology still appears not 
sufficiently settled for investing major development efforts.

\medskip
So the front-ends of Isabelle and of \sisac{} go opposite directions: the former 
towards a sophisticated workbench for engineers running on a workstation and the 
latter towards an easy-to-use tool running on a tablet or even on a mobile. 
However, the current state of Isabelle/jEdit appears a good compromise to start 
adopting it as front-end for an educational tool in introductory courses at engineering faculties.

The issues in adopting Isabelle/jEdit as front-end for \sisac{} are challenging: 
Differences on the side of the math-engine have been discussed in 
\S\ref{ssec:isab-isac}. Even more challenging is a controversial 
architecture: Isabelle is a heavy tool for one engineer with, in an optimal 
case, a server farm running automated provers in parallel --- while \sisac{} is 
a client in the cloud, for participants of courses served by a central session 
management (see \S\ref{ssec:user-guide}). The yellow area in the middle of 
Fig.\ref{fig:ast-trans} on p.\pageref{fig:ast-trans}, the interfaces between 
front-ends and math-engines raises serious questions to be tackled in 
cooperation with the Isabelle team.

\subsection{Towards a Professional Tool}\label{ssec:professional}
Students' projects (more than 30 in 
number\footnote{\url{http://www.ist.tugraz.at/isac/Credits}}
) were appropriate for experimenting 
with ideas and technologies. But \sisac's code base has become too complex 
(45957 lines of code (LOC) production code, 18465 LOC test code for the front-end, 33573 LOC 
production code, 42384 LOC test code for the mathematics-engine) and the 
planned development tasks are demanding such that the next phase of development 
cannot be carried out without full-time software professionals. The following tasks have 
to be accomplished towards a professional tool for engineering education at 
universities:
\medskip
\begin{compactenum}
\item Shift \sisac's programming language to Isabelle's function package 
according to \S\ref{ssec:lucin}.
\item Adapt Isabelle's document model to the needs of \sisac{}: introduce 
session management in order to allow for user guidance (\S\ref{ssec:user-guide}), adapt parsers 
to \sisac's calculations (instead of Isar proofs), adapt the interface of 
\sisac's math-engine to PIDE's document model. Adoption of this model solves 
several problems, e.g. explicit theory imports in the specification phase 
(presently blocked by libisabelle as mentioned in \S\ref{ssec:isab-isac}).
\item Improve the graphical formula editor; clarify location of ast-translations 
according to Fig.\ref{fig:ast-trans} and embed the editor into Isabelle/jEdit.
\item Provide authoring tools:
  \begin{compactenum}
  \item for dialogue authoring according to \S\ref{ssec:user-guide}
  \item for adding explanations in HTML~5.0 to math knowledge
   according to \S\ref{ssec:knowl}
  \item for debugging programs using the Lucas-Interpreter, see 
\S\ref{ssec:lucin}
  \end{compactenum}
\item Implement a graph-based tool for the specification phase as shown in
Fig.\ref{fig:sub-probl}. Consider partial evaluation of physical units only
according to~\cite[p.27]{isac-doc2}.
\item Replace the only non-open-source library~\cite{mkienl-bakk} in \sisac{} by 
proprietary code using Scala's \texttt{match}. Further shifts from Java to Scala 
can be expected (the editor~\cite{mmahringer} is already written in Scala).
\item Implement tools for collaboration as envisaged already by \sisac's 
architecture: tutors remotely inspect students work, lecturers demonstrate a 
calculation on students' computers, students construct solutions collaboratively 
according to \S\ref{ssec:user-guide}.
\item Make \sisac{} water-proof for written exams (where the system's help 
functionality and transparency are reduced to ``exam mode'' by the 
\texttt{DialogGuide} \S\ref{ssec:user-guide}).
\item Extend \sisac's calculations in a way that they can hold incorrect steps.
Unlike proofs, where only incomplete steps are handled (by \texttt{oops} in 
Isabelle), for students also incorrect steps might be 
instructive: What can happen when I integrate over poles, or when I allow 
division by zero?
\end{compactenum}
\medskip
The efforts for implementing the above points can be estimated with ten man-years.

\section{Conclusions}\label{sec:conclusion}
After almost two decades of prototyping is clear, where the \sisac{} software and 
respective innovations in teaching and learning are most helpful: in mathematics 
education at technical faculties. The original conception
provided a ``complete'', transparent and interactive model of mathematics 
(\S\ref{ssec:self-expl}, \S\ref{ssec:user-guide}). As 
such it is appreciated most by academic lecturers, who have respective expertise 
and freedom for innovation; here \sisac's ability to connect basic courses to 
advanced labs as well as to connect intuitive application 
with abstract mathematics come into effect 
most usefully and most effectively.

Long-term prototyping definitely paid off; the user requirements evolved over 
time and time could not have been shorter for finding out, what users (students 
and teachers) can request from a new generation of systems based on TP 
technology. A recent requirements analysis in cooperation with universities of 
applied science in Upper Austria confirmed the original conception  and 
contributed novel ideas (\S\ref{ssec:spec-phase}). Experience 
with various technologies settled, the system interfaces stabilised. 
Tasks for further integrating \sisac's math-engine (\S\ref{sec:math-eng}) 
into Isabelle are clarified.
As 
mentioned in \S\ref{ssec:professional}, the original design stood the test and 
the code quality remained such that improving the prototype towards a 
professional system appears more efficient than re-implementing from scratch.

The variety of contributions of more than thirty students not only balanced the 
system, this work also involved various kinds of expertise from the respective 
advisors. This way a network of interdisciplinary expertise arose, a perfect 
requisite to constitute competent developer teams for large development projects. The tasks 
to be performed in order to develop a professional tool are well defined and the 
required efforts can be reliably estimated (\S\ref{ssec:professional}).

\medskip
The estimated effort of ten man-years, however, is difficult to finance. So it is 
good to know that the \sisac-project is a free-rider on the successfully ongoing propagation of 
formal methods, including TP, which is enforced by increasing complexity in 
technology as addressed by ``internet of things'', ``industry 4.0", ``systems of 
systems'' and the like.

Once formalisations of engineering disciplines are elaborated, \sisac{} can add 
interactive problem solving with little effort; and who knows, ``systems that 
explain themselves'' might even be useful for knowledge management in 
development units of specific companies.

\section*{Acknowledgements}
The remarkable origins of the \sisac{}project were described in \S\ref{sec:intro}. The continuation is supported by hosting code repositories and administrative data at Graz University of Technology, the institutes IST and IICM and at Johannes Kepler University Linz, the institutes RISC and IIS. The evolvement of the project was and is driven by interdisciplinary expertise, where particularly warm thanks go to the advisors of theses within \sisac{} (orderd by time of first involvement): Franz Wotawa, Clemens Heuberger, Stephan Dreiseitl, Denis Helic, Bernhard Aichernig, Christian G\"utl, Jens Knoop and Wolfgang Schreiner.

\nocite{*}
\bibliographystyle{eptcs}
\bibliography{references}
\end{document}